\pdfoutput=1
\documentclass[11pt,a4paper]{article}

\usepackage[utf8]{inputenc}
\usepackage[T1]{fontenc}
\usepackage{lmodern}
\usepackage[margin=2.5cm]{geometry}
\usepackage{amsmath,amssymb}
\usepackage{graphicx}
\usepackage{authblk}
\usepackage[numbers,sort&compress]{natbib}
\usepackage[colorlinks=true,allcolors=blue]{hyperref}

\title{Multiple harmonics from a hand-struck tube: a simple and precise measurement of the speed of sound}

\author[1]{Francisco Echegorri}
\author[2]{Mart\'in Monteiro}
\author[3]{Cecilia Stari}
\author[3,*]{Arturo C. Mart\'i}
\affil[1]{IPA, ANEP, Montevideo, Uruguay}
\affil[2]{Universidad ORT Uruguay, Montevideo, Uruguay}
\affil[3]{Instituto de F\'isica, Universidad de la Rep\'ublica, Montevideo, Uruguay}
\affil[*]{Corresponding author: \texttt{marti@fisica.edu.uy}}

\date{}

\begin{document}

\maketitle

\begin{abstract}
The speed of sound in air can be determined from the resonance
frequencies of the harmonic waves inside a thin cylindrical tube. A
pressure wave is generated in the tube by striking several times one of its ends
with an open hand and keeping the hand in place. In this way pulses
of short duration are produced which propagate along the tube and
decompose into a superposition of harmonic waves whose frequencies
are determined by the harmonics present in a tube open at one end and
closed at the other. The tube thus acts as a filter that reinforces
those harmonic waves exhibiting an acoustic pressure antinode (particle displacement node) at the closed end and an acoustic pressure node (particle displacement antinode) at the open end. Their frequencies are determined
by these boundary conditions, while waves with different frequencies
cancel out by destructive interference. Recording the sound with a
microphone connected to a computer and analyzing its spectrum with
freely available software, the resonance frequencies are obtained and
the speed of sound is determined with an accuracy better than
$0.5\%$. The whole experiment requires nothing but a tube, a microphone
and free software, so it is inexpensive, takes a few minutes to set up
and is well suited to introductory physics laboratories.
\end{abstract}

\section{\label{Intro}Introduction}

Sound waves in air consist of small variations of the air pressure
with respect to its equilibrium value, $p(x,t)=P(x,t)-P_0$, called the
acoustic pressure. These variations propagate as a wave in the
medium. The motion of the air particles is collinear with the
direction of propagation, so these are longitudinal waves. For waves
in one dimension the acoustic pressure satisfies the wave
equation~\cite{kinsler2000}
\begin{equation}
\frac{\partial^{2}p}{\partial t^{2}}
= c^{2}\,\frac{\partial^{2}p}{\partial x^{2}},
\label{eq:wave}
\end{equation}
where $c=\sqrt{B/\rho}$ is the speed of sound, $B$ is the adiabatic
bulk modulus and $\rho$ is the density of air. In the approximation of
air as an ideal gas, the speed of sound is given by
$c=\sqrt{\gamma R T/\overline{M}}$, where $\gamma$ is the ratio of the
specific heats at constant pressure and constant volume, $T$ is the
absolute temperature (in kelvin) and $\overline{M}$ is the mean molar mass of
air. At $0\,^{\circ}$C, $c=331.4$~m/s, and at a temperature $T$ it can be approximated by~\cite{kinsler2000}
\begin{equation}
c = 331.4 + 0.60\, T ,
\label{eq:temperature}
\end{equation}
where, in this case, $T$ is expressed in degrees Celsius.
The determination of the speed of sound is a classical experiment
in introductory physics courses, and in recent years numerous low-cost
variants based on computers, smartphones and everyday materials have
been
proposed~\cite{gil2014,hellesund2019,monteiro2023,monteiro2018,vogt2012,monteiro2015measuring,Hirth2015,kasper2015stationary,coban2020,gomez2014}.
Different approaches have been reported, including measurements based
on the time delay between two microphones~\cite{gomez2014}, resonance
methods using open or closed tubes~\cite{Hirth2015,kasper2015stationary},
techniques involving smartphone microphones and dedicated
applications~\cite{vogt2012}, and electronic
circuits that measure the propagation time of
sound~\cite{coban2020}. These experiments provide accessible
alternatives to traditional laboratory setups, but they generally rely
on the measurement of a single quantity, either a time interval or one
resonance frequency, so that their precision is limited to a few
percent.

In this context, we propose a simple and precise alternative based on
the frequency-domain analysis of the sound recorded inside a PVC tube
struck with an open hand, using only a microphone, a personal computer
and free audio-analysis software. 
Building upon the impulse excitation method described in our previous work~\cite{monteiro2023}, where a single hand clap was employed to probe resonant air columns, a single impulsive excitation
simultaneously excites the whole harmonic series of the tube, so that a
large number of resonances can be identified in the Fast Fourier
Transform (FFT) of one recording. Since the speed of sound is obtained
from the slope of a straight line fitted to many harmonics at once,
rather than from an individual time interval or a single resonance
frequency, the determination is strongly constrained and the method is
considerably more precise and robust than comparable low-cost
experiments.

In an \emph{open--closed} tube the acoustic pressure has an antinode (particle displacement node) at the closed end and a node (particle displacement antinode) slightly beyond the open end, at a
distance $\delta$ from the rim. This end correction accounts for the
radiation of sound from the open end into the surrounding air.
Textbooks give $\delta \approx 0.6\,a$, with $a$ the radius of the
tube~\cite{kinsler2000}; we use the more precise low-frequency value
$\delta = 0.6133\,a$ derived by Levine and Schwinger for an unflanged
circular pipe~\cite{levine1948radiation}. The normal modes of the tube
then have frequencies at odd multiples of the fundamental,
\begin{equation}
f_n = \frac{(2n-1)\,c}{4 L'}, \qquad n=1,2,3,\dots
\label{eq:harmonics}
\end{equation}
where
\begin{equation}
L' = L + \delta = L + 0.6133\, a
\label{eq:effective}
\end{equation}
is the effective acoustic length of a tube of geometric length $L$.

\section{\label{Methods}Materials and methods}

The experiment requires only inexpensive and readily available
materials: a PVC tube (one or more tubes of different lengths and
diameters can be used), a microphone connected to the sound card or a
USB port of a personal computer, a measuring tape and a thermometer.
On the software side, the sound is recorded and its spectrum computed
with the free, open-source application Audacity~\cite{audacity}. The resonance frequencies can be obtained directly from the spectrum or determined by fitting the peaks using data-analysis software such as Logger Pro, Scilab, Python, or similar tools. The subsequent linear fit can be performed using any suitable data-analysis software or spreadsheet.

\begin{figure}
\centering
\includegraphics[width=0.48\textwidth]{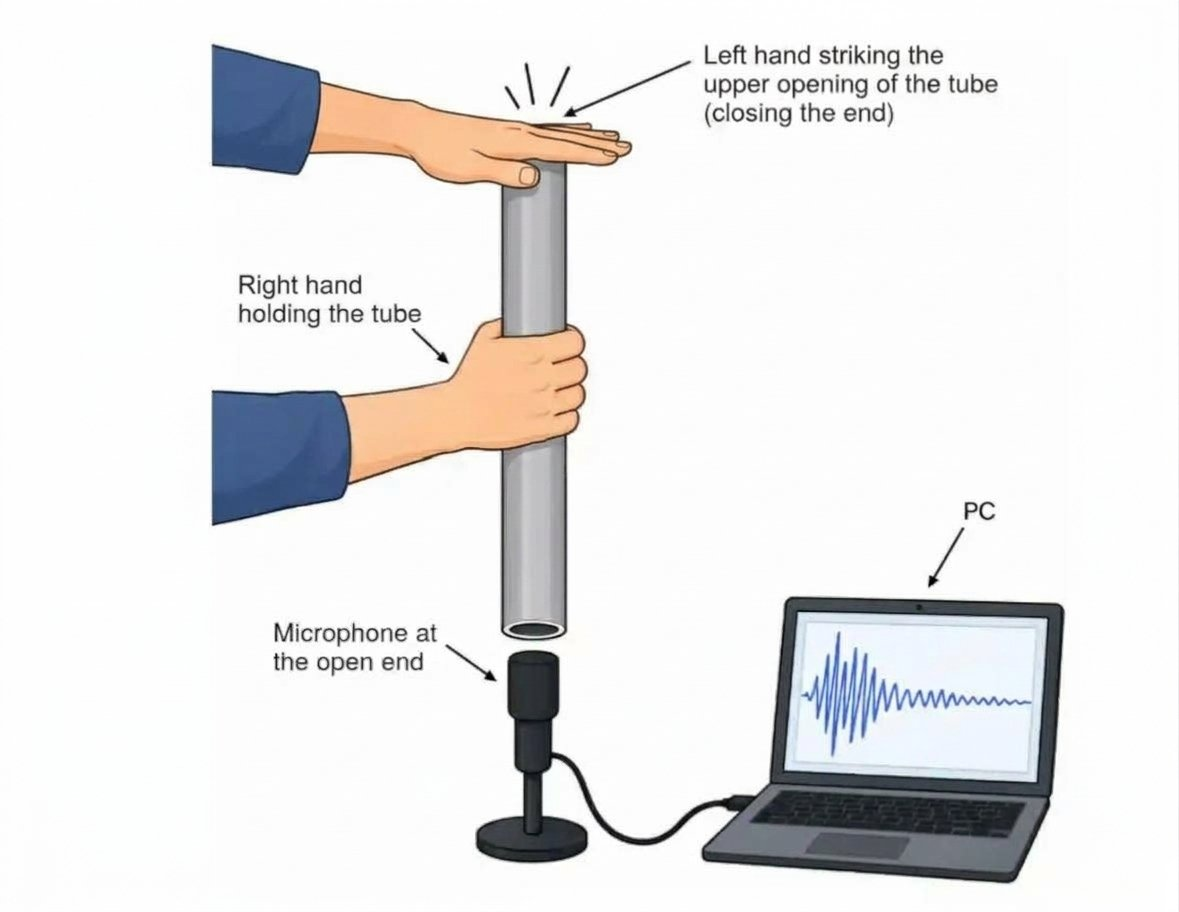}
\caption{Experimental setup.}
\label{fig:tube}
\end{figure}

To generate the acoustic signal, one end of the tube is repeatedly struck with an
open hand and the hand is kept in place, as shown in
figure~\ref{fig:tube}. In this way the tube behaves as a tube closed
at one end (the hand) and open at the other, and the short pressure
pulse excites the normal modes given by
equation~(\ref{eq:harmonics}).
This excitation procedure follows the impulse approach introduced in our previous study~\cite{monteiro2023}, where hand percussions were used to excite standing waves in resonant columns. From an acoustic perspective, clapping the palm firmly against the tube rim produces a short pressure pulse, whose duration is much smaller than the period of the fundamental mode. Such a short pulse has a broad and nearly flat spectrum, so that all the normal modes of the tube are excited simultaneously with comparable amplitudes. The air column then acts as a sharp acoustic filter through successive reflections between the boundaries. Since each transient signal persists for approximately $\Delta t \approx 0.4~\mathrm{ s}$ before decaying below the background noise, the acoustic wave travels an overall distance of $\approx 130 ~\mathrm{ m}$. In a tube of length $L \approx 1.5~\mathrm{ m}$ ($2L \approx 3.0~\mathrm{ m}$ per round trip), the pulse undergoes more than 40 round trips. This repeated propagation selectively reinforces the harmonic components that satisfy the boundary conditions (a pressure antinode at the closed end and a node at the open mouth) through constructive interference, while off-resonance frequencies are suppressed by destructive interference. Crucially, keeping the palm firmly seated against the rim throughout the impact is necessary to preserve the closed-end boundary condition; any premature bounce would alter the acoustic impedance during the transient ring-down.

We have carried out this experiment with students on repeated occasions. In general, all of them manage to produce a sequence of strikes that faithfully reproduces the spectrum of the open--closed tube. We recommend striking the rim with the centre of the palm and keeping the hand in place during the strike. A good strike is felt as a vibration, or resonance, in the hand and is heard in the same way, as a resonant sound rather than a dry slap. In our experience, this activity also brings a relaxed touch to the laboratory session.

The resonance frequencies are determined with a microphone connected
to the computer and the application Audacity. Upon opening Audacity, a
sampling rate must be selected in the preferences according to the
frequency range to be observed in the fast Fourier transform
(FFT). For a PVC tube about one metre long or longer, a sampling rate
of 6000~Hz is sufficient, giving an FFT range of 3000~Hz that allows a
large number of harmonics to be observed and measured. The sampling
rate is set in the menu Edit~$\rightarrow$~Preferences, selecting
6000~Hz as the project sampling rate.
The experiment can also be performed using built-in microphones in laptops or smartphones. Laptop microphones generally have reduced sensitivity below 100~Hz, while smartphone microphones may detect frequencies down to approximately 50~Hz, depending on the model. In either case, the speed of sound can be determined from the higher harmonics, even if the lowest-frequency peaks are not detected.

\begin{figure}
\centering
\includegraphics[width=0.9\textwidth]{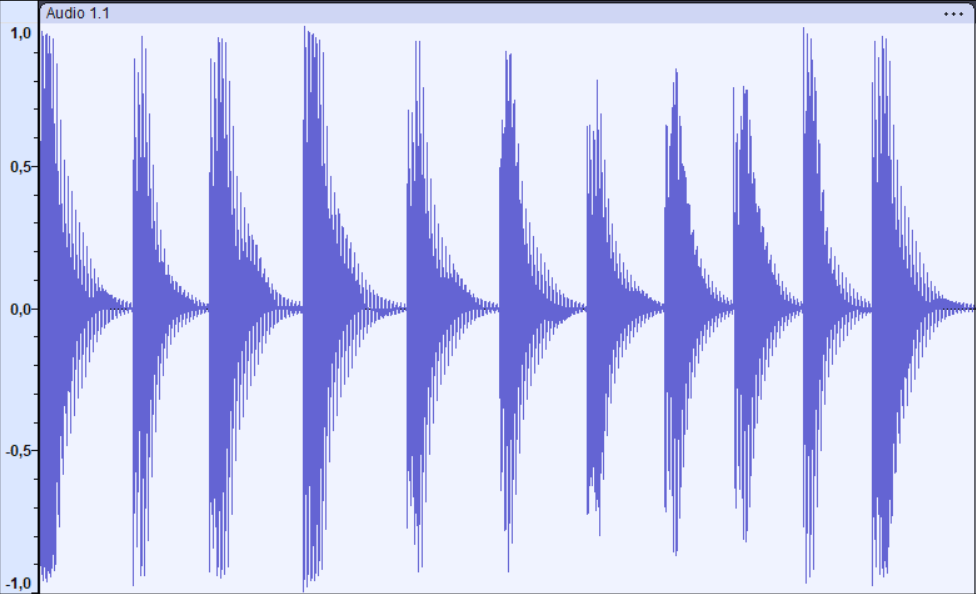}
\caption{Audacity snapshot of the raw recording, showing the series of
pulses produced by successive strikes on one end of the tube.}
\label{fig:signal}
\end{figure}

Figure~\ref{fig:signal} shows the raw recording: each strike is
followed by a rapidly decaying oscillation and by a silent
interval. In the measurement reported here 11 strikes were recorded,
each lasting approximately 0.36~s. 

\begin{figure}
\centering
\includegraphics[width=0.9\textwidth]{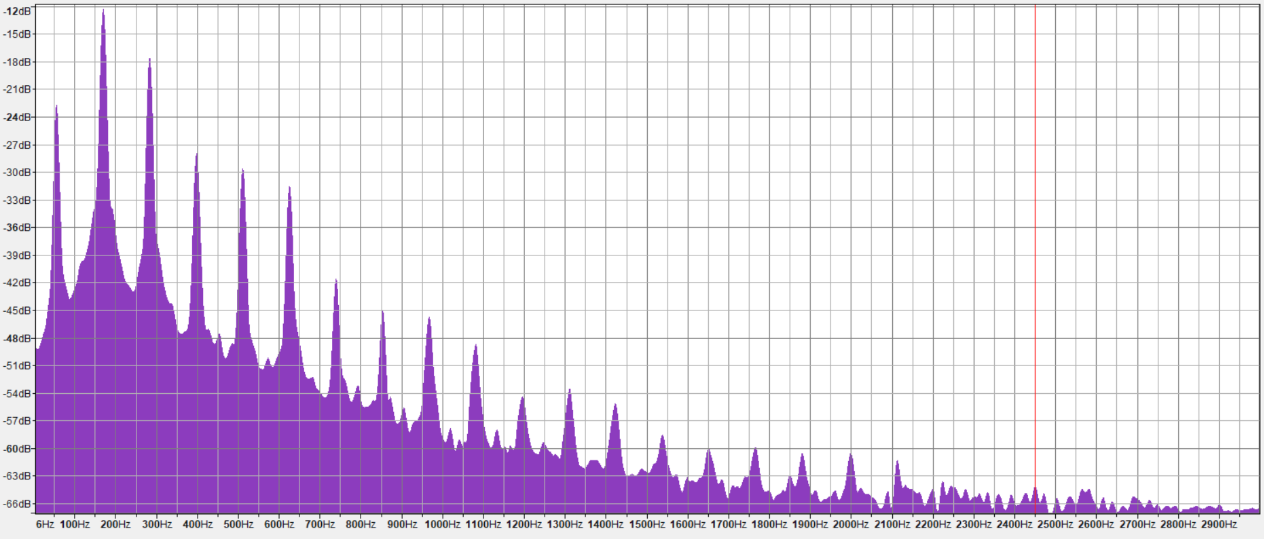}
\caption{Frequency spectrum of the recorded signal obtained with
Audacity. The resonance peaks, corresponding to the odd harmonics of
an open--closed tube, are clearly visible.}
\label{fig:spectrum}
\end{figure}

Finally, the acoustic signal is selected and the spectrum is obtained
in the menu Analyze~$\rightarrow$~Plot Spectrum
(figure~\ref{fig:spectrum}). The FFT size is chosen as $2^{N}$, $N$
being related to the number of samples, and a linear frequency scale
is selected for the horizontal axis. If the frequencies are to be
measured with higher precision, the FFT can be exported as a text file
containing one column with the frequency in Hz and one column with the
sound pressure level in dB. 

\begin{figure}
    \centering
    \includegraphics[width=0.9\linewidth]{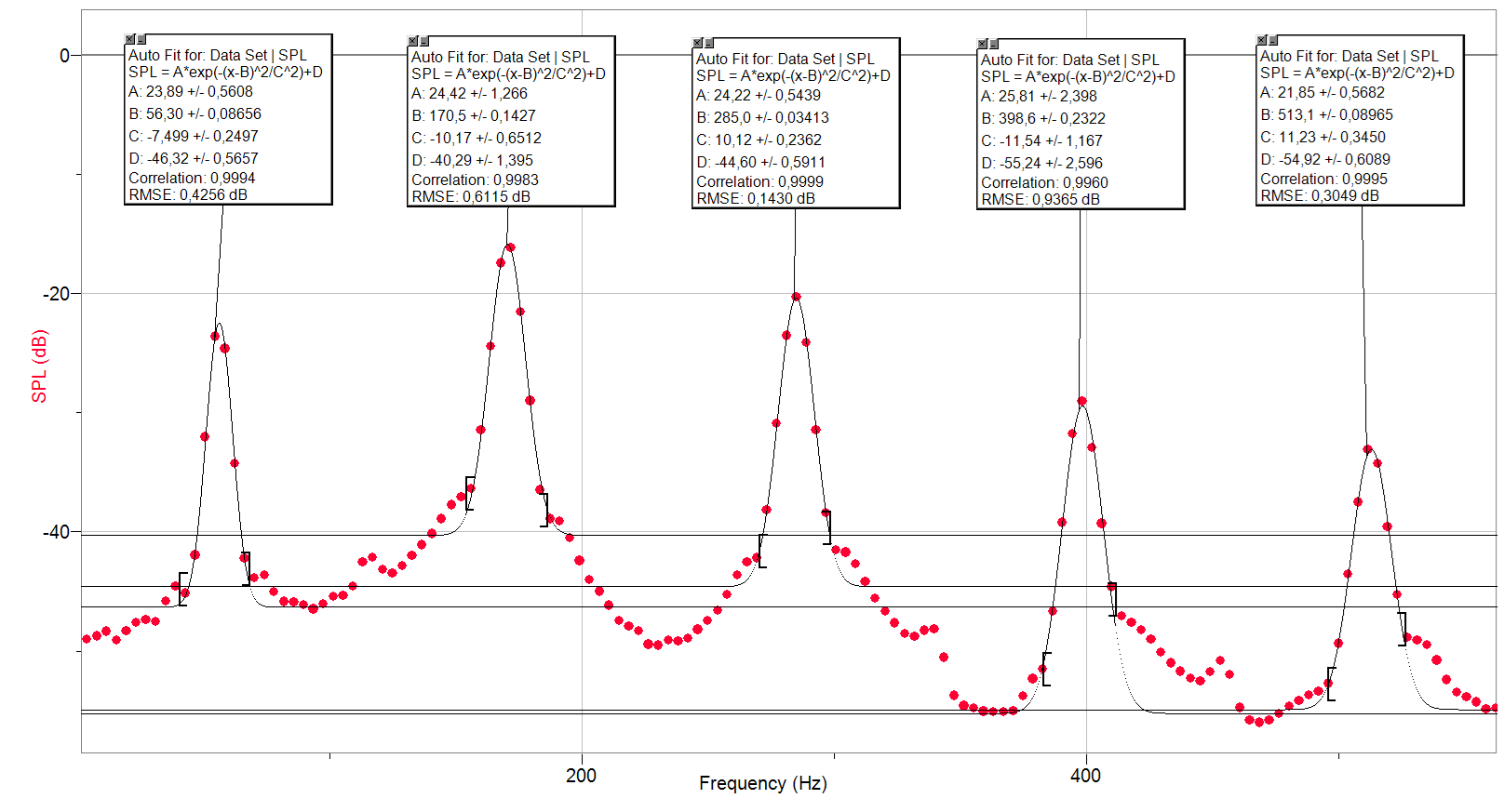}
    \caption{Gaussian fits to the first five resonance peaks of the exported spectrum. The vertical brackets indicate the 7 to 9 data points around each apex selected for the fit to exclude the background noise floor. Note that standard software outputs often display an excessive, unaligned number of digits by default; for subsequent analysis, the centroid frequencies (B) and their standard uncertainties are properly rounded to one decimal place (e.g., $f_1 = 56.3 \pm 0.1\,\mathrm{Hz}$, $f_2 = 170.5 \pm 0.1\,\mathrm{Hz}$, $f_3 = 285.0 \pm 0.1\,\mathrm{Hz}$, $f_4 = 398.6 \pm 0.2\,\mathrm{Hz}$, and $f_5 = 513.1 \pm 0.1\,\mathrm{Hz}$).}
    \label{fig:gaussians}
\end{figure}

At the introductory level, students can read the resonance frequencies directly by placing the cursor at the apex of each peak in the Audacity spectrum, without further analysis, provided care is taken to ensure that the frequency resolution $f_s / N$ is smaller than $1\,\mathrm{Hz}$ (for instance, with $f_s = 6000\,\mathrm{Hz}$ and an FFT size of $N = 8192$, the frequency spacing between consecutive bins is $\Delta f = f_s / N = 0.73\,\mathrm{Hz}$). Alternatively, the exported data columns can be imported into data-analysis software such as Logger Pro, Scilab, Python, or similar tools to fit the peaks. As an example, figure~\ref{fig:gaussians} shows Gaussian fits to the first resonance peaks, from which the peak frequencies are obtained with uncertainties of a fraction of a hertz. When using automated curve-fitting tools, instructors should remind students to round the raw software output so that the uncertainty is expressed with one (or at most two) significant figures and the measured value matches the same decimal place.
Once the resonance frequencies have been determined, either directly from the spectrum or through Gaussian fitting, they can be plotted against the harmonic numbers and fitted with a straight line using any suitable data-analysis software or a spreadsheet.

\section{\label{Results}Example of a measurement}

For this experiment a PVC tube of length $L=1.498(1)$~m and radius
$a=1.85(5)$~cm was used. Substituting these values into
equation~(\ref{eq:effective}) gives an effective length
$L' = 1.5093(13)$~m.

\begin{figure}
\centering
\includegraphics[width=0.9\textwidth]{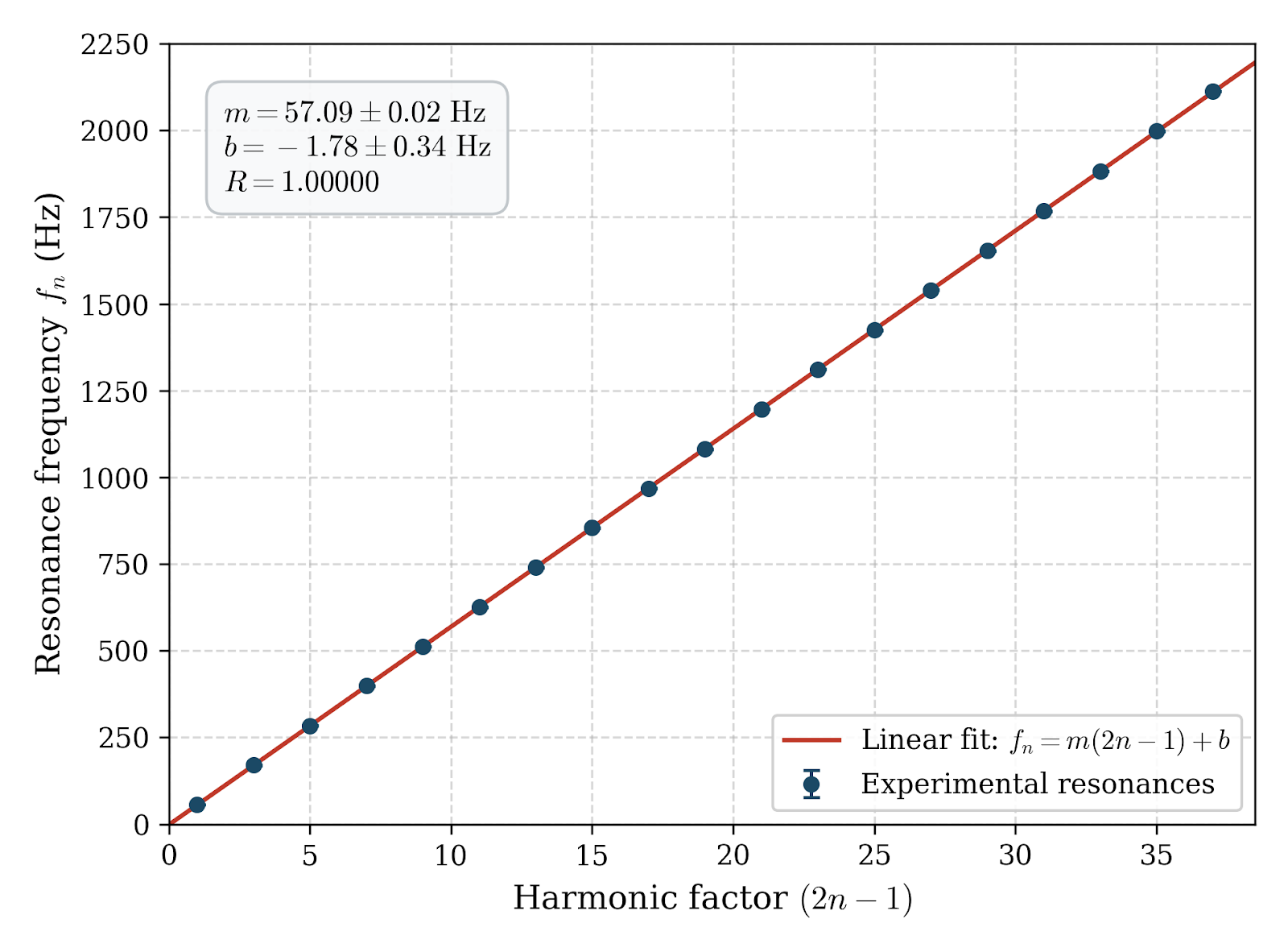}
\caption{Resonance frequency as a function of $2n-1$ and least-squares
linear fit. The slope, $c/4L'=57.09(2)$~Hz, yields the speed of sound.}
\label{fig:fit}
\end{figure}

A table of the harmonic number $n$ and the corresponding resonance
frequency is built, together with a column with the odd numbers
$2n-1$. According to equation~(\ref{eq:harmonics}), the frequency as a
function of $2n-1$ must be a straight line through the origin with
slope $c/4L'$:
\begin{equation}
f_n = \frac{c}{4L'}\,(2n-1).
\label{eq:line}
\end{equation}
Figure~\ref{fig:fit} shows the resonance frequency as a function of
$2n-1$ for the 19 harmonics identified in the spectrum, together with
the least-squares linear fit, which yields
\begin{equation}
\frac{c}{4L'} = 57.09 \pm 0.02~\mathrm{Hz} ,
\end{equation}
with a correlation coefficient of 1.0000. The speed of sound is then
\begin{equation}
c_{exp} = 4 \times 1.5093(13) \times 57.09(2) = 344.7(3)~\mathrm{m/s}.
\end{equation}

The experimental result can be compared with the accepted value for the laboratory's specific ambient conditions, obtained via an online calculator~\cite{sengpiel}. For a temperature of
$19.2(5)\,^{\circ}$C, an atmospheric pressure of 101.325~kPa and a
relative humidity of 65(5)\%, then reference value is
$c_{ref}=343.7(6)$~m/s. The absolute difference between the experimental and reference values is $\delta c = 1.0$~m/s, which corresponds to a minor relative discrepancy of:
\begin{equation}
\frac{\delta c}{c}\times 100 = 0.30\%
\end{equation}
Alternatively, using the simpler approximation from equation~(\ref{eq:temperature}) (which neglects the effect of humidity), yields $c_{dry}=331.4+0.60\times 19.2(5)=342.9(6)$~m/s. Even when omitting humidity, this simplified estimate remains remarkably close to the empirical measurement, within a $0.5\%$ margin of error.

\section{\label{Conclusion}Conclusion}

We have presented a simple, inexpensive and accurate experiment to
determine the speed of sound in air from the resonance frequencies of
a cylindrical tube open at one end and closed at the other. The
experiment requires only a PVC tube, a microphone, a personal computer
and free software, and can be carried out in any introductory physics
laboratory or even at home, in line with other recent low-cost
proposals~\cite{hellesund2019,monteiro2023,monteiro2018,vogt2012}. The
possibility of measuring a large number of harmonics (19 in the
example presented) and of fitting them globally with a straight line
makes the method remarkably precise: the value obtained differs from
the accepted one by only $0.30\%$. Beyond the determination of the
speed of sound, the experiment allows students to work with key
concepts such as standing waves, boundary conditions, harmonics, the
end correction of open pipes and Fourier analysis of a real signal.

\bibliographystyle{unsrtnat}
\bibliography{sound_speed_resonance}

\end{document}